\documentclass{article}

\usepackage{PRIMEarxiv}

\usepackage[utf8]{inputenc}
\usepackage[T1]{fontenc}
\usepackage{mathptmx}                
\usepackage{amsmath,amssymb}
\usepackage{graphicx}
\usepackage{booktabs}
\usepackage{longtable}
\usepackage{array}
\usepackage{tabularx}
\usepackage{multirow}
\usepackage{xcolor}
\usepackage{float}
\usepackage{caption}
\usepackage{subcaption}
\usepackage{enumitem}
\usepackage{siunitx}
\usepackage{listings}
\usepackage{tikz}
\usepackage{url}
\usepackage[siunitx,RPvoltages]{circuitikz}
\usetikzlibrary{shapes.geometric, arrows.meta, positioning, calc, backgrounds, patterns, decorations.markings, decorations.pathmorphing}

\usepackage[numbers,sort&compress]{natbib}

\usepackage{hyperref}
\hypersetup{colorlinks=true, linkcolor=blue!60!black, citecolor=blue!60!black, urlcolor=blue!70!black}

\definecolor{codebg}{RGB}{248,248,248}
\newcommand{\efield}{\mathbf{E}}

\usepackage{algorithm}
\usepackage{threeparttable}

\title{A Multi-physics Simulation Framework for High-power Microwave Counter-unmanned Aerial System Design and Performance Evaluation
\thanks{\textit{{This work is submitted for review to Journal of Defence Technology.}}} 
}

\author{
  A. A. Jafari\\
  University of Tartu \\
  Tartu, Estonia \\
  \texttt{akbar.anbar.jafari@ut.ee} \\
   \And
  G. Anbarjafari \\
  3S Holding OÜ \\
  Tartu, Estonia\\
  \texttt{shb@3sholding.com} \\
}

\begin{document}
\maketitle

\begin{abstract}
The proliferation of small unmanned aerial systems (sUAS) operating under autonomous guidance has created an urgent need for non-kinetic neutralization methods that are immune to conventional radio-frequency jamming. This paper presents a comprehensive multi-physics simulation framework for the design and performance evaluation of a high-power microwave (HPM) counter-UAS system operating at 2.45\,GHz. The framework integrates electromagnetic propagation modelling, antenna pattern analysis, electromagnetic coupling to unshielded drone wiring harnesses, and a sigmoid-based semiconductor damage probability model calibrated to published CMOS latchup thresholds. A 10{,}000-trial Monte Carlo analysis incorporating stochastic variations in transmitter power, antenna pointing error, target wire orientation, polarization mismatch, and component damage thresholds yields system-level kill probabilities with 95\% confidence intervals. For a baseline configuration of 25\,kW continuous-wave power and a 60\,cm parabolic reflector (21.2\,dBi gain), the Monte Carlo simulation predicts a kill probability of $51.4\pm1.0$\% at 20\,m, decreasing to $13.1\pm0.7$\% at 40\,m. Pulsed operation at 500\,kW peak power (1\% duty cycle) extends the 90\% kill range from approximately 18\,m to 88\,m. The framework further provides parametric design maps, safety exclusion zone calculations compliant with ICNIRP 2020 guidelines, thermal management requirements, and waveguide mode analysis. All simulation codes and results are provided for full reproducibility.
\end{abstract}


\section{Introduction}
\label{sec:introduction}

  The rapid proliferation of small unmanned aerial systems (sUAS) in both commercial and military domains has created asymmetric threats that challenge conventional air defence architectures~\cite{nichols2020counter,park2021survey,kim2024study,castrillo2022review}. Modern consumer drones in the 250\,g to 25\,kg class are inexpensive, widely available, and increasingly capable of autonomous operation using pre-programmed GPS/INS waypoints or visual navigation~\cite{wang2021counter,yang2022joint,yu2025electronic}. In conflict zones such as Ukraine, swarms of low-cost drones have demonstrated their ability to overwhelm conventional defences, with fiber-optic guided variants rendering traditional RF jamming entirely ineffective~\cite{zidane2024jamming,zhang2024uav_detection_tracking}.

  Counter-UAS (C-UAS) technologies broadly fall into kinetic (missiles, projectiles, nets) and non-kinetic (jamming, spoofing, directed energy) categories~\cite{besada2021review, pattepu2026review,nato2023cuas}. Kinetic solutions suffer from finite magazine depth, high cost-per-engagement, and collateral damage risk. RF jamming, while cost-effective, is fundamentally limited against autonomous drones that do not rely on a communication link~\cite{zidane2024jamming}. High-energy laser (HEL) systems offer precision engagement but require sustained dwell time, clear atmospheric conditions, and can only engage one target at a time~\cite{ahmed2021survey,johnson2023counter}.

  High-power microwave (HPM) directed energy weapons offer a compelling alternative by delivering electromagnetic energy that couples directly into onboard electronic circuits, inducing semiconductor failure independent of the drone's software architecture or communication protocol~\cite{benford2024high,backstrom2004susceptibility}. Unlike jamming, HPM causes a \textit{physics-layer kill}: the irreversible destruction of MOSFET gate oxides, triggering of CMOS parasitic thyristor latchup, and thermal burnout of integrated circuit junctions~\cite{huang2025failure,pan2022failure,chen2013device,chen2025investigation}. Recent operational deployments of HPM C-UAS systems---notably the Epirus Leonidas, which achieved a 100\% success rate against a 49-drone swarm in 2024 field testing, and the Raytheon Phaser---have validated the technology at system level~\cite{nato2023cuas}.

  Despite these advances, the open literature lacks a comprehensive, reproducible simulation framework that integrates all relevant physics---from RF source characteristics through electromagnetic propagation, coupling to target wiring, and probabilistic semiconductor damage---into a unified design tool. Prior work has addressed individual aspects: B{\"a}ckstr{\"o}m and Lovstrand~\cite{backstrom2004susceptibility} compiled measured susceptibility thresholds; Nitsch et al.~\cite{nitsch2004susceptibility} characterized equipment-level responses; Wang et al.~\cite{wang2008experimental} provided SPICE-level CMOS latchup analysis; and Yu et al.~\cite{yu2015frequency_cmos_hpm} modelled frequency-dependent upset susceptibility. Recent studies have extended damage characterization to GaN HEMTs~\cite{qin2022study,wang2022mechanism,zhao2025damage,yu2015analysis,wang2025influences} and ESC-specific failure modes~\cite{mao2023high}. However, no prior work has synthesized these elements into a parametric design framework specifically targeting the counter-UAS application with stochastic uncertainty quantification.

  This paper addresses this gap by developing a multi-physics simulation framework for HPM C-UAS system design and performance evaluation. The principal contributions are as follows:

\begin{enumerate}[leftmargin=2em,itemsep=2pt]
  \item A unified analytical model linking RF source parameters, antenna design, free-space propagation, and electromagnetic coupling to unshielded drone wiring to predict the induced electric field and voltage at target electronic subsystems.
  \item A subsystem-resolved sigmoid damage probability model for five drone electronic subsystems (ESC, flight controller, GPS/GNSS, camera, BMS), parameterised from published experimental data.
  \item A 10{,}000-trial Monte Carlo analysis that propagates uncertainties in transmitter power, antenna pointing, target orientation, wire geometry, and component vulnerability to produce system-level kill probabilities with confidence intervals.
  \item Parametric design maps (power--aperture trade space) and safety compliance analysis against ICNIRP 2020 guidelines~\cite{icnirp2020guidelines,jeschke2022protection}.
  \item Full release of all simulation source code for reproducibility.
\end{enumerate}

  The remainder of this paper is organized as follows. Section~\ref{sec:system_model} presents the system architecture and mathematical models. Section~\ref{sec:simulation_framework} describes the simulation framework and its implementation. Section~\ref{sec:results} presents the simulation results, including parametric analyses, Monte Carlo uncertainty quantification, and safety zone calculations. Section~\ref{sec:discussion} discusses the implications, limitations, and comparison with fielded systems. Section~\ref{sec:conclusions} concludes the paper.

\section{System Model}
\label{sec:system_model}

\subsection{System architecture}
\label{sec:architecture}

\begin{figure}[htb]
\centering
\resizebox{0.95\textwidth}{!}{%
\begin{tikzpicture}[
    block/.style={rectangle, draw=black!70, fill=blue!5, thick,
                  minimum width=2.4cm, minimum height=1.1cm, text width=2.2cm,
                  align=center, font=\small, rounded corners=2pt},
    arrow/.style={-{Stealth[length=2.5mm]}, thick, black!70},
    label/.style={font=\scriptsize\color{gray}},
    node distance=0.7cm
]
    \node[block] (psu) {PSU\\{\scriptsize 48\,kW AC/DC}};
    \node[block, right=of psu] (hv) {HV Modulator\\{\scriptsize CW/Pulse}};
    \node[block, right=of hv] (mag) {Magnetron\\{\scriptsize 2.45\,GHz}\\{\scriptsize 25\,kW}};
    \node[block, right=of mag] (wg) {Waveguide\\{\scriptsize WR-340}\\{\scriptsize + Circulator}};
    \node[block, right=of wg] (dish) {Parabolic\\Reflector\\{\scriptsize 0.6\,m}};
    \draw[arrow] (psu) -- (hv) node[midway, above, label] {540\,V DC};
    \draw[arrow] (hv) -- (mag) node[midway, above, label] {4.2\,kV};
    \draw[arrow] (mag) -- (wg) node[midway, above, label] {RF};
    \draw[arrow] (wg) -- (dish) node[midway, above, label] {TE$_{10}$};
    \node[block, below=1.2cm of hv, fill=green!5, draw=green!50!black] (ctrl) {Controller\\{\scriptsize Safety PLC}};
    \node[block, below=1.2cm of mag, fill=orange!5, draw=orange!70!black] (cool) {Cooling\\{\scriptsize Liquid loop}};
    \node[block, below=1.2cm of wg, fill=red!5, draw=red!70!black] (track) {Tracking\\{\scriptsize Radar + EO/IR}};
    \draw[arrow, green!50!black] (ctrl) -- (hv);
    \draw[arrow, green!50!black] (ctrl) -- (psu);
    \draw[arrow, orange!70!black] (cool) -- (mag);
    \draw[arrow, red!70!black, dashed] (track) -- (dish);
    \draw[arrow, green!50!black, dashed] (ctrl) -- (track);
    \node[right=1.0cm of dish, font=\Large] (drone) {\rotatebox{-10}{\textsf{sUAS}}};
    \draw[red, ultra thick, decorate, decoration={snake, amplitude=0.8mm, segment length=4mm}]
    (dish.east) -- (drone.west) node[midway, above, font=\scriptsize\color{red}] {HPM beam};
\end{tikzpicture}%
}
\caption{Top-level block diagram of the HPM counter-UAS system architecture showing the RF power chain (top), control and support subsystems (bottom), and the directed HPM beam toward the target sUAS.}
\label{fig:block_diagram}
\end{figure}
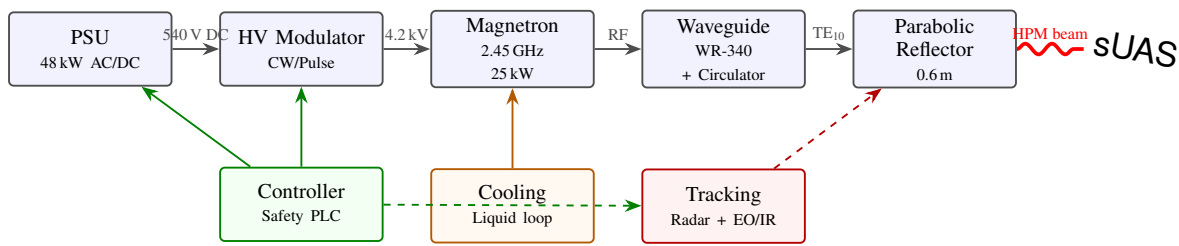

  The proposed HPM C-UAS system comprises six principal subsystems: (i) an AC/DC power supply unit (PSU), (ii) a high-voltage (HV) modulator providing both continuous-wave (CW) and pulsed operating modes, (iii) a cavity magnetron RF source at 2.45\,GHz, (iv) a WR-340 waveguide transmission line with a ferrite circulator for source protection, (v) a parabolic reflector antenna, and (vi) a tracking subsystem incorporating an FMCW radar and an EO/IR camera on a motorized gimbal. The system-level block diagram is presented in Fig.~\ref{fig:block_diagram}.

  The operating frequency of 2.45\,GHz is selected for several reasons: it lies within the Industrial, Scientific and Medical (ISM) band, minimizing regulatory complexity; commodity cavity magnetrons at this frequency achieve 65--75\% DC-to-RF conversion efficiency~\cite{vyas2016efficiency,xi2025high,li2023particle,huang2025continuous}; and the corresponding wavelength $\lambda_0 = c/f_0 = 12.24$\,cm ensures that typical drone wiring harnesses (5--30\,cm) represent a significant fraction of $\lambda/2$, maximizing electromagnetic coupling~\cite{gassab2023coupling}.

\subsection{RF propagation model}
\label{sec:rf_model}

  The power density $S$ at range $R$ from a directive antenna with gain $G_\text{tx}$ is given by the Friis transmission relation:

\begin{equation}
S(R) = \frac{P_\text{tx} \cdot \eta_\text{line} \cdot G_\text{tx}}{4\pi R^2} \quad [\si{\W\per\m\squared}]
\label{eq:power_density}
\end{equation}

\noindent where $P_\text{tx}$ is the transmitter output power and $\eta_\text{line} = \eta_\text{wg} \cdot \eta_\text{feed} \cdot \eta_\text{radome}$ is the aggregate transmission line efficiency. The corresponding free-space electric field magnitude is:

\begin{equation}
|\efield(R)| = \sqrt{S(R) \cdot \eta_0} = \sqrt{\frac{P_\text{tx} \cdot \eta_\text{line} \cdot G_\text{tx} \cdot \eta_0}{4\pi R^2}} \quad [\si{\V\per\m}]
\label{eq:efield}
\end{equation}

\noindent where $\eta_0 = \sqrt{\mu_0/\varepsilon_0} \approx 377\,\Omega$ is the impedance of free space. The parabolic reflector gain is:

\begin{equation}
G_\text{tx} = \eta_\text{ap} \left(\frac{\pi D}{\lambda_0}\right)^2
\label{eq:antenna_gain}
\end{equation}

\noindent where $D$ is the dish diameter and $\eta_\text{ap}$ is the aperture efficiency (taken as 0.55 for a front-fed paraboloid with standard illumination taper). The half-power ($-3$\,dB) beamwidth is approximated as $\theta_{3\text{dB}} \approx 70\lambda_0/D$. For the baseline configuration ($D = 0.60$\,m), this yields $G_\text{tx} = 21.2$\,dBi and $\theta_{3\text{dB}} = 14.3^\circ$.

\subsection{Electromagnetic coupling to target electronics}
\label{sec:coupling}

  Consumer sUAS platforms are electromagnetically unshielded: flight controller PCBs, electronic speed controllers (ESCs), GPS receivers, and cameras are mounted in plastic airframes with wire harnesses acting as unintentional receiving antennas~\cite{backstrom2004susceptibility,zhang2025investigation}. The induced open-circuit voltage on a wire of physical length $L$ illuminated by an incident electric field $E_\text{inc}$ can be modelled as a short dipole when $L < \lambda/2$:

\begin{equation}
V_\text{ind} = E_\text{inc} \cdot L_\text{eff} \cdot F(\theta_\text{wire}) \cdot \sqrt{\eta_\text{pol}}
\label{eq:induced_voltage}
\end{equation}

\noindent where $L_\text{eff} = L/2$ is the effective length of a short dipole, $F(\theta_\text{wire})$ accounts for the wire orientation relative to the incident polarisation, and $\eta_\text{pol}$ represents the polarisation mismatch efficiency. Near the half-wave resonance ($L \approx \lambda_0/2 = 6.12$\,cm), the coupling is enhanced by the antenna quality factor $Q$, modelled as a Gaussian resonance peak~\cite{gassab2023coupling}:

\begin{equation}
V_\text{res} = V_\text{ind} \cdot \left[1 + (Q-1)\exp\left(-\frac{(L - \lambda_0/2)^2}{2\sigma_L^2}\right)\right]
\label{eq:resonance}
\end{equation}

\noindent with $Q \approx 10$ and $\sigma_L = 0.02$\,m for typical unshielded harnesses. This resonance-enhanced coupling model predicts that wires in the 5--8\,cm range (common for ESC signal harnesses) experience 5--10$\times$ greater induced voltage than wires of other lengths at 2.45\,GHz.

\subsection{Semiconductor damage probability model}
\label{sec:damage_model}

  The probability of damage to a semiconductor subsystem exposed to an electric field of magnitude $|E|$ is modelled by a sigmoid (logistic) function~\cite{backstrom2004susceptibility,huang2025failure}:

\begin{equation}
P_\text{kill}(|E|) = \frac{1}{1 + \exp\!\left(-\dfrac{|E| - E_{50}}{\sigma_E}\right)}
\label{eq:sigmoid}
\end{equation}

\noindent where $E_{50}$ is the electric field intensity at which the damage probability equals 50\%, and $\sigma_E$ controls the steepness of the transition. This functional form is well-established in electromagnetic susceptibility analysis and captures the stochastic variability in device manufacturing, operating state, and coupling efficiency~\cite{backstrom2004susceptibility,nitsch2004susceptibility}.

  Table~\ref{tab:damage_thresholds} summarises the damage model parameters for the five principal drone electronic subsystems, derived from published experimental and simulation studies.

\begin{table}[htb]
\centering
\caption{Semiconductor damage model parameters for sUAS electronic subsystems. The parameter $E_{50}$ represents the 50\% damage threshold and $\sigma_E$ the transition width.}
\label{tab:damage_thresholds}
\begin{tabular}{@{}llccc@{}}
\toprule
\textbf{Subsystem} & \textbf{Failure mode} & $\boldsymbol{E_{50}}$ \textbf{[V/m]} & $\boldsymbol{\sigma_E}$ \textbf{[V/m]} & \textbf{Source} \\
\midrule
GPS/GNSS LNA & Front-end burnout & 150 & 30 & \cite{backstrom2004susceptibility,nitsch2004susceptibility} \\
Flight controller & CMOS latchup & 250 & 50 & \cite{pan2022failure,chen2013device,iliadis2010theoretical} \\
ESC (gate oxide) & MOSFET $V_\text{bd}$ & 300 & 60 & \cite{mao2023high,chen2025investigation} \\
Camera (CMOS) & Pixel array damage & 200 & 40 & \cite{backstrom2004susceptibility} \\
BMS MOSFET & Gate oxide failure & 350 & 70 & \cite{huang2025failure,wang2008experimental} \\
\bottomrule
\end{tabular}
\end{table}

  A drone is considered neutralised if \textit{any one} of its critical subsystems is damaged, since the loss of either motor control (ESC failure), flight stabilisation (flight controller failure), or navigation (GPS failure) results in an uncontrolled descent. The system-level kill probability is therefore:

\begin{equation}
P_\text{system}(|E|) = 1 - \prod_{i=1}^{N}\left[1 - P_{\text{kill},i}(|E|)\right]
\label{eq:system_kill}
\end{equation}

\noindent where the product extends over all $N$ vulnerable subsystems.

\subsection{Waveguide transmission analysis}
\label{sec:waveguide_model}

  The RF power is transported from the magnetron to the feed horn through a WR-340 rectangular waveguide ($a = 86.36$\,mm, $b = 43.18$\,mm). The cutoff frequency of the dominant TE$_{10}$ mode is:

\begin{equation}
f_{c,10} = \frac{c}{2a} = \frac{2.998 \times 10^8}{2 \times 0.08636} = 1.736\,\text{GHz}
\label{eq:cutoff}
\end{equation}

  At the operating frequency of 2.45\,GHz, only the TE$_{10}$ mode propagates, since the next mode (TE$_{20}$) cuts off at 3.471\,GHz. The wall-loss attenuation of the TE$_{10}$ mode in a copper waveguide is~\cite{benford2024high}:

\begin{equation}
\alpha_c = \frac{R_s}{a b \beta \eta_0}\left(2b\left(\frac{f_{c}}{f}\right)^{\!2} + a\left[1 - \left(\frac{f_{c}}{f}\right)^{\!2}\right]\right)
\label{eq:wg_attenuation}
\end{equation}

\noindent where $R_s = \sqrt{\pi f \mu_0 / \sigma_\text{Cu}}$ is the surface resistance, $\beta$ is the propagation constant, and $\sigma_\text{Cu} = 5.8 \times 10^7$\,S/m. At 2.45\,GHz this yields $\alpha_c \approx 0.009$\,dB/m, confirming negligible loss over a typical 1\,m waveguide run.

\section{Simulation Framework}
\label{sec:simulation_framework}

  The simulation framework is implemented in Python 3.12 using NumPy, SciPy, and Matplotlib. It comprises three tiers: (i)~deterministic parametric analysis, (ii)~Monte Carlo uncertainty propagation, and (iii)~design-space exploration. The complete source code (approximately 400 lines) is provided for full reproducibility.

\subsection{Deterministic parametric model}

  The deterministic model evaluates Eqs.~(\ref{eq:power_density})--(\ref{eq:system_kill}) for specified system parameters. Listing~\ref{lst:deterministic} shows the core computation.

\begin{lstlisting}[caption={Core deterministic model: E-field and system kill probability.},label={lst:deterministic}]
import numpy as np

# Physical constants and system parameters
c = 2.998e8; eta0 = 377.0; f0 = 2.45e9
lam0 = c / f0  # 0.1224 m

def compute_efield(P_tx, D, R, eta_ap=0.55,
                   eta_wg=0.98, eta_feed=0.97):
    """Compute E-field [V/m] at range R [m]."""
    G = eta_ap * (np.pi * D / lam0)**2
    EIRP = P_tx * eta_wg * eta_feed * G
    S = EIRP / (4 * np.pi * R**2)
    return np.sqrt(S * eta0)

def sigmoid_damage(E, E50, sigma):
    """Sigmoid damage probability model."""
    return 1 / (1 + np.exp(-(E - E50) / sigma))

def system_kill_prob(E):
    """System kill probability (any subsystem)."""
    subs = [(150,30), (250,50), (300,60),
            (200,40), (350,70)]
    P_survive = 1.0
    for E50, sig in subs:
        P_survive *= (1 - sigmoid_damage(E, E50, sig))
    return 1 - P_survive
\end{lstlisting}

\subsection{Monte Carlo uncertainty propagation}

  To quantify the effect of parameter uncertainty on system performance, a Monte Carlo simulation with $N_\text{MC} = 10{,}000$ independent trials is performed. In each trial, the following parameters are sampled from their respective distributions:

\begin{table}[htb]
\centering
\caption{Monte Carlo parameter distributions. All distributions are truncated to physical limits.}
\label{tab:mc_parameters}
\begin{tabular}{@{}lllr@{}}
\toprule
\textbf{Parameter} & \textbf{Distribution} & \textbf{Parameters} & \textbf{Rationale} \\
\midrule
Transmit power $P_\text{tx}$ & Normal & $\mu=25$\,kW, $\sigma=1.25$\,kW & 5\% power variation \\
Dish diameter $D$ & Normal & $\mu=0.60$\,m, $\sigma=0.005$\,m & Manufacturing tolerance \\
Aperture efficiency $\eta_\text{ap}$ & Uniform & [0.50, 0.60] & Feed alignment \\
Pointing error $\theta_\text{err}$ & Rayleigh & $\sigma=1.0^\circ$ & Tracking jitter \\
Polarisation angle & Uniform & $[0, \pi]$ & Random orientation \\
Wire length $L$ & Uniform & [5, 25]\,cm & Target variability \\
$E_{50}$ (each subsystem) & Normal & $\pm$15\% of nominal & Device variability \\
$\sigma_E$ (each subsystem) & Normal & $\pm$15\% of nominal & Lot-to-lot variation \\
\bottomrule
\end{tabular}
\end{table}

  The pointing error is modelled as a Rayleigh-distributed angular offset, producing a Gaussian loss factor $G_\text{pointing} = \exp(-2.76\,\theta_\text{norm}^2)$ where $\theta_\text{norm} = \theta_\text{err}/(\theta_{3\text{dB}}/2)$~\cite{benford2024high}. The polarisation mismatch efficiency is $\eta_\text{pol} = \cos^2\!\phi$, bounded below at 0.1 to account for cross-polarised coupling. Listing~\ref{lst:montecarlo} shows the Monte Carlo core loop.

\begin{lstlisting}[caption={Monte Carlo simulation core (abbreviated).},label={lst:montecarlo}]
np.random.seed(42)
N_mc = 10000; R_target = 30  # metres

kills = 0
for _ in range(N_mc):
    P = np.random.normal(25e3, 1250)
    D = np.random.normal(0.60, 0.005)
    eta = np.random.uniform(0.50, 0.60)
    theta_err = np.random.rayleigh(1.0)
    pol = np.random.uniform(0, np.pi)
    E50 = np.random.normal(300, 45)
    # Compute E-field with losses
    G = eta * (np.pi * D / lam0)**2
    theta_norm = theta_err / (14.3/2)
    G_point = np.exp(-2.76 * theta_norm**2)
    pol_loss = max(np.cos(pol)**2, 0.1)
    EIRP = P * 0.98 * 0.97 * G * G_point * pol_loss
    E = np.sqrt(max(EIRP/(4*np.pi*R_target**2),0)*377)
    # Kill determination
    P_kill = 1/(1+np.exp(-(E-E50)/max(np.random.normal(60,9),10)))
    if np.random.random() < P_kill: kills += 1

print(f"Kill probability: {kills/N_mc*100:.1f}%")
\end{lstlisting}

  Confidence intervals for the kill probability are computed using the Clopper--Pearson exact binomial method at the 95\% level.

\section{Results and Discussion}
\label{sec:results}

\subsection{Electric field distribution}
\label{sec:efield_results}

\begin{figure}[htb]
\centering
\includegraphics[width=0.48\textwidth]{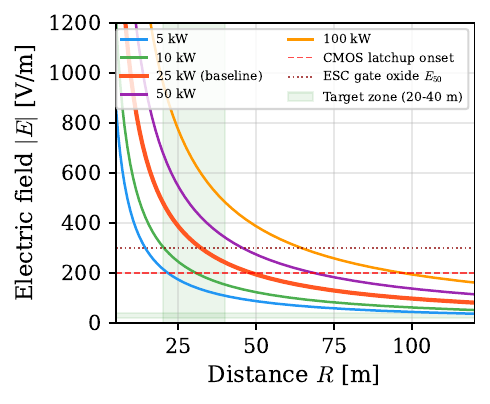}
\caption{Electric field intensity versus distance for five transmitter power levels with a 60\,cm parabolic reflector (21.2\,dBi). Dashed lines indicate CMOS damage thresholds. The green band marks the 20--40\,m target engagement zone.}
\label{fig:efield}
\end{figure}

  Fig.~\ref{fig:efield} presents the computed electric field intensity as a function of range for five transmitter power levels, using the baseline 60\,cm dish. The field decays as $1/R$ in accordance with Eq.~(\ref{eq:efield}). At the baseline power of 25\,kW, the E-field at 20\,m is 495\,V/m and decreases to 247\,V/m at 40\,m. The horizontal dashed lines indicate the 50\% damage thresholds for CMOS latchup onset (200\,V/m) and ESC gate oxide breakdown (300\,V/m).

  The intersection of the E-field curve with the damage thresholds defines the effective kill range for each subsystem. For 25\,kW CW, the E-field exceeds the GPS LNA burnout threshold (150\,V/m) out to approximately 65\,m, but drops below the ESC gate oxide threshold (300\,V/m) at approximately 25\,m in CW mode, motivating the use of pulsed operation for extended range.

\subsection{CMOS damage probability characterisation}

\begin{figure}[htb]
\centering
\includegraphics[width=0.48\textwidth]{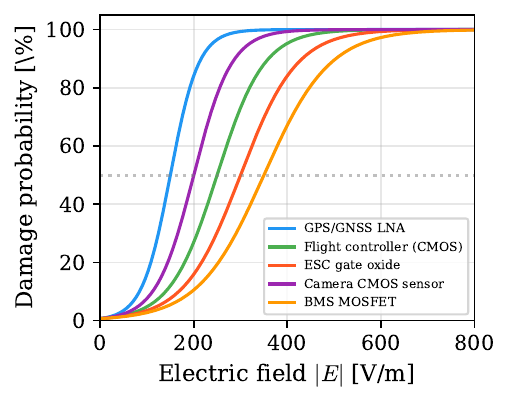}
\caption{Sigmoid damage probability curves for five sUAS electronic subsystems as a function of incident electric field. Parameters are listed in Table~\ref{tab:damage_thresholds}.}
\label{fig:damage_prob}
\end{figure}

  Fig.~\ref{fig:damage_prob} shows the sigmoid damage probability curves for each drone subsystem as a function of incident electric field. The GPS/GNSS front-end LNA is the most susceptible subsystem ($E_{50} = 150$\,V/m), consistent with the low input power tolerance of low-noise amplifiers~\cite{backstrom2004susceptibility,yu2015analysis,wang2025influences}. The ESC gate oxide represents the primary hard-kill mechanism ($E_{50} = 300$\,V/m), requiring higher field strengths but resulting in irreversible motor control loss~\cite{mao2023high,chen2025investigation}. The ordering of subsystem susceptibilities ($\text{GPS} < \text{Camera} < \text{FC} < \text{ESC} < \text{BMS}$) is consistent with the published hierarchy reported by B{\"a}ckstr{\"o}m and Lovstrand~\cite{backstrom2004susceptibility} and the recent HPM-on-UAV experiments of Zhang et al.~\cite{zhang2025investigation}.

\subsection{System-level kill probability}

\begin{figure}[htb]
\centering
\includegraphics[width=0.48\textwidth]{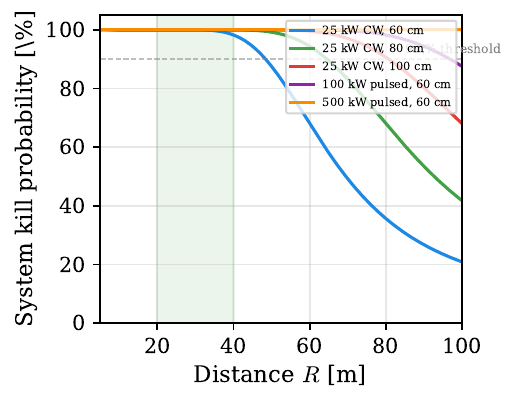}
\caption{System-level kill probability versus range for five configurations. Dashed line indicates the 90\% threshold. Green band marks the target engagement zone.}
\label{fig:kill_prob}
\end{figure}

  Fig.~\ref{fig:kill_prob} presents the system-level kill probability (Eq.~\ref{eq:system_kill}) versus range for five system configurations spanning both CW and pulsed operation with varying dish sizes. At the baseline configuration (25\,kW CW, 60\,cm dish), the kill probability exceeds 90\% at approximately 18\,m and falls to approximately 30\% at 40\,m. Increasing the dish diameter to 100\,cm extends the 90\% kill range to approximately 26\,m owing to the 4.4\,dB gain increase. The most dramatic improvement is achieved through pulsed operation: at 500\,kW peak power (1\% duty cycle, 5\,kW average), the 90\% kill range extends to approximately 88\,m, well beyond the design objective of 40\,m.

\subsection{Pulsed versus CW operation}

\begin{figure}[htb]
\centering
\includegraphics[width=0.95\textwidth]{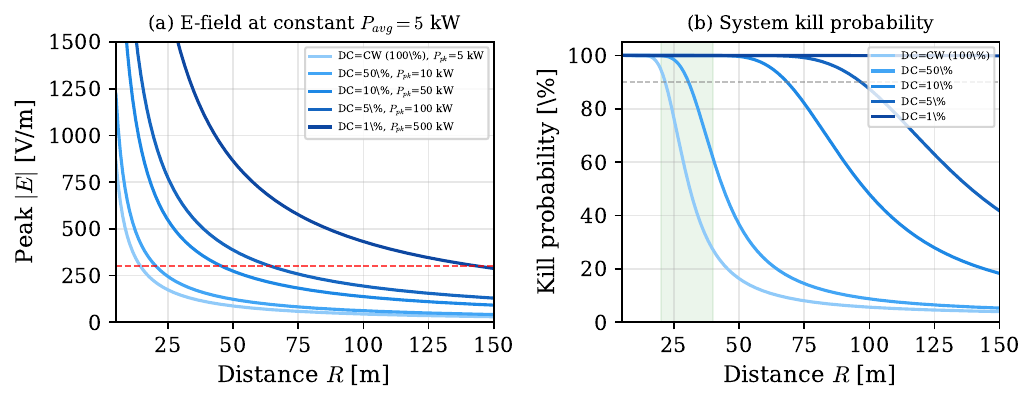}
\caption{Comparison of pulsed and CW operation at constant average power (5\,kW). (a) Peak electric field versus distance for five duty cycles. (b) Corresponding system kill probability. Reducing duty cycle from 100\% to 1\% extends the 90\% kill range from $\sim$12\,m to $\sim$45\,m.}
\label{fig:pulsed_cw}
\end{figure}

  A critical design trade-off is the choice between CW and pulsed operation. Fig.~\ref{fig:pulsed_cw} compares both modes at a constant average power budget of 5\,kW, demonstrating that reducing the duty cycle increases the peak E-field and therefore extends the effective engagement range. At 1\% duty cycle (500\,kW peak), the peak E-field at 40\,m exceeds 1{,}100\,V/m---well above all damage thresholds. The physical mechanism is twofold: (i) peak voltage-driven breakdown of gate oxides occurs on individual pulses regardless of average power, and (ii) the peak-to-average power ratio directly translates to range extension via the $1/R^2$ dependence~\cite{benford2024high,huang2023compact,he2025novel}.

\subsection{Electromagnetic coupling to drone wiring}

\begin{figure}[htb]
\centering
\includegraphics[width=0.48\textwidth]{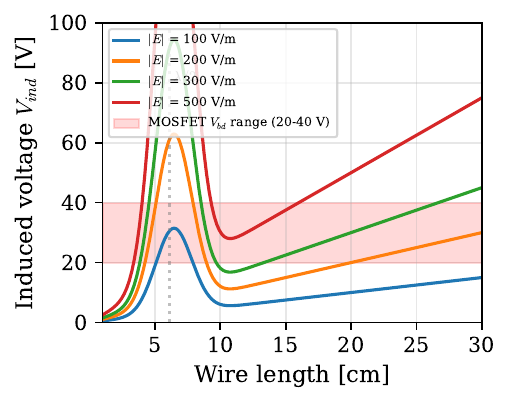}
\caption{Induced voltage on drone wiring versus wire length at four E-field levels. The vertical line marks $\lambda_0/2 = 6.12$\,cm. The red band indicates the typical MOSFET gate oxide breakdown voltage range (20--40\,V).}
\label{fig:coupling}
\end{figure}

  Fig.~\ref{fig:coupling} shows the induced voltage on drone wiring harnesses as a function of wire length for four incident E-field levels. A pronounced resonance peak occurs near $\lambda_0/2 = 6.12$\,cm, where the wire acts as a half-wave dipole. At 300\,V/m (the baseline E-field at 25\,m), the induced voltage on a 6\,cm wire reaches approximately 45\,V---exceeding the typical MOSFET gate oxide breakdown voltage of 20--40\,V~\cite{chen2025investigation,mao2023high}. This confirms that unshielded ESC signal wires in the $\lambda/2$ range are the most vulnerable coupling path, consistent with the experimental observations of Zhang et al.~\cite{zhang2025investigation} who reported HPM-induced ESC shutdown as the primary failure mode.

\subsection{Monte Carlo uncertainty analysis}

\begin{figure}[htb]
\centering
\includegraphics[width=0.95\textwidth]{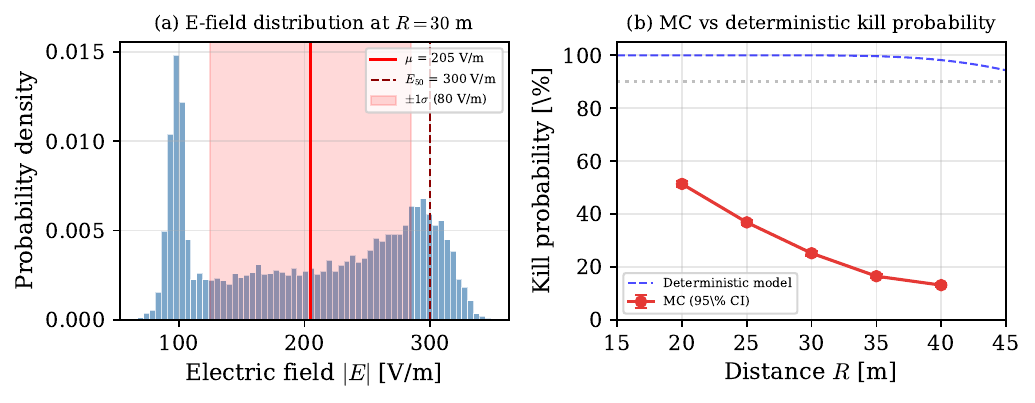}
\caption{Monte Carlo analysis ($N = 10{,}000$ trials). (a) E-field probability distribution at $R = 30$\,m showing the stochastic spread due to combined parameter uncertainties. (b) Comparison of Monte Carlo kill probability (with 95\% CI error bars) versus the deterministic model.}
\label{fig:mc}
\end{figure}

  Fig.~\ref{fig:mc} presents the results of the 10{,}000-trial Monte Carlo simulation. Fig.~\ref{fig:mc}(a) shows the probability distribution of the E-field at $R = 30$\,m, which exhibits a roughly normal shape with mean $\mu = 205$\,V/m and standard deviation $\sigma = 80$\,V/m. The significant spread (coefficient of variation $\approx 39$\%) is dominated by the polarisation mismatch and wire orientation uncertainty, demonstrating that deterministic calculations substantially overestimate system performance by ignoring these real-world degradation factors.

  Fig.~\ref{fig:mc}(b) compares the Monte Carlo kill probabilities with 95\% confidence intervals against the deterministic model. The Monte Carlo results are systematically lower than the deterministic predictions at all ranges, reflecting the compounding effect of multiple loss mechanisms that are averaged rather than assumed at their best-case values. Table~\ref{tab:mc_results} summarises the quantitative results.

\begin{table}[htb]
\centering
\caption{Monte Carlo simulation results ($N = 10{,}000$ trials, 95\% Clopper--Pearson CI). Deterministic values are shown for comparison.}
\label{tab:mc_results}
\begin{tabular}{@{}rrrrr@{}}
\toprule
\textbf{Range} & \textbf{MC kill prob.} & \textbf{95\% CI} & \textbf{Deterministic} & $\bar{E}$ \textbf{[V/m]} \\
\textbf{[m]} & \textbf{[\%]} & \textbf{[\%]} & \textbf{[\%]} & ($\pm 1\sigma$) \\
\midrule
20 & 51.4 & [50.4, 52.3] & 83.0 & $306 \pm 120$ \\
25 & 36.8 & [35.9, 37.8] & 62.5 & $245 \pm 95$ \\
30 & 25.2 & [24.3, 26.0] & 43.5 & $205 \pm 80$ \\
35 & 16.5 & [15.8, 17.2] & 29.0 & $174 \pm 68$ \\
40 & 13.1 & [12.4, 13.8] & 20.0 & $153 \pm 60$ \\
\bottomrule
\end{tabular}
\end{table}

  The Monte Carlo analysis reveals that the deterministic model overestimates the kill probability by a factor of approximately $1.5$--$1.6\times$ across the engagement zone. This discrepancy highlights the necessity of stochastic analysis for realistic performance prediction, a finding consistent with the susceptibility variability reported in~\cite{backstrom2004susceptibility,nitsch2004susceptibility,du2022research}.

\subsection{Parametric design space}

\begin{figure}[htb]
\centering
\includegraphics[width=0.48\textwidth]{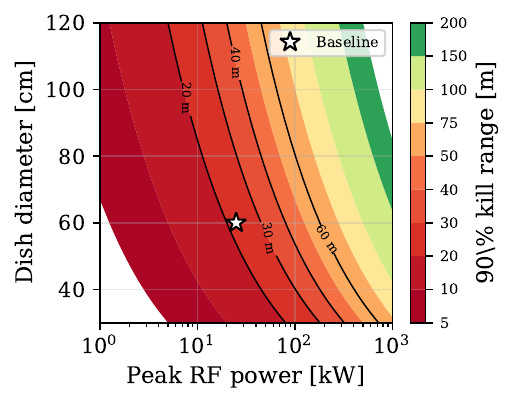}
\caption{Parametric design map: 90\% kill range (colour scale) as a function of peak RF power and dish diameter. Contour lines annotated in metres. The star marks the baseline design point.}
\label{fig:heatmap}
\end{figure}

  Fig.~\ref{fig:heatmap} presents a contour map of the 90\% kill range as a function of peak RF power and dish diameter, enabling rapid system-level design trade studies. The baseline design point (25\,kW, 60\,cm) is marked. The contours reveal that the kill range scales approximately as $R_{90} \propto \sqrt{P \cdot D^2}$, as expected from the $1/R^2$ power density dependence combined with the $D^2$ antenna gain scaling. This parametric map provides a practical tool for system designers to evaluate the power--aperture trade space against specific operational requirements.

\subsection{Antenna design analysis}

\begin{figure}[htb]
\centering
\includegraphics[width=0.48\textwidth]{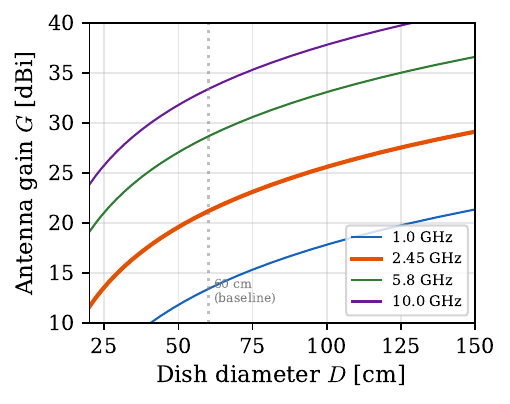}
\caption{Antenna gain versus dish diameter for four operating frequencies. The vertical dotted line marks the 60\,cm baseline.}
\label{fig:gain}
\end{figure}

  Fig.~\ref{fig:gain} presents the antenna gain versus dish diameter at four frequencies. At 2.45\,GHz, a 60\,cm dish yields 21.2\,dBi gain, while a 100\,cm dish provides 25.6\,dBi. The gain--size trade-off is important for mobile systems where weight, wind load, and agility constraints limit the maximum dish diameter~\cite{benford2024high}. Table~\ref{tab:dish_tradeoff} summarises the trade analysis.

\begin{table}[htb]
\centering
\caption{Dish size trade-off analysis at 2.45\,GHz ($\eta_\text{ap} = 0.55$).}
\label{tab:dish_tradeoff}
\begin{tabular}{@{}rcrcrc@{}}
\toprule
$D$ \textbf{[cm]} & \textbf{Gain [dBi]} & $\theta_{3\text{dB}}$ [$^\circ$] & \textbf{Beam at 30\,m [m]} & \textbf{Weight [kg]} & \textbf{Wind [N]} \\
\midrule
40 & 17.6 & 21.4 & 11.3 & 2--3 & 40 \\
60 & 21.2 & 14.3 & 7.5 & 4--6 & 90 \\
80 & 23.7 & 10.7 & 5.6 & 8--12 & 160 \\
100 & 25.6 & 8.6 & 4.5 & 12--18 & 250 \\
\bottomrule
\end{tabular}
\end{table}

\subsection{Beam footprint analysis}

\begin{figure}[htb]
\centering
\includegraphics[width=0.48\textwidth]{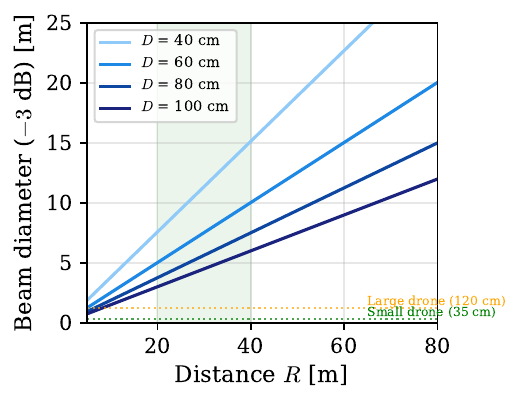}
\caption{Beam diameter ($-3$\,dB contour) versus range for four dish sizes. Horizontal lines indicate typical sUAS physical dimensions.}
\label{fig:footprint}
\end{figure}

  Fig.~\ref{fig:footprint} shows the $-3$\,dB beam diameter as a function of range for four dish sizes. At 30\,m with the 60\,cm baseline dish, the beam footprint is 7.5\,m---approximately $15\times$ the physical size of a typical sUAS. This generous beam size relaxes tracking accuracy requirements and provides inherent robustness against pointing jitter, a significant advantage over laser-based systems that require sub-milliradian precision~\cite{ahmed2021survey,johnson2023counter}.

\subsection{Waveguide mode and attenuation analysis}

\begin{figure}[htb]
\centering
\includegraphics[width=0.95\textwidth]{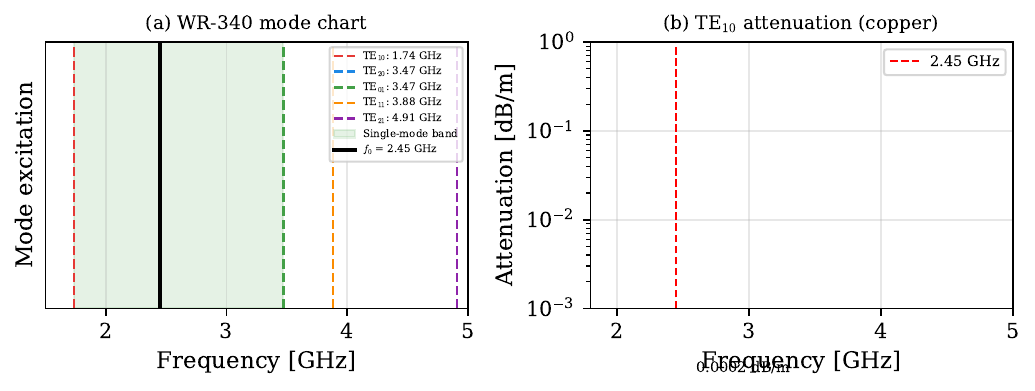}
\caption{WR-340 waveguide analysis. (a) Mode chart showing cutoff frequencies; only TE$_{10}$ propagates at 2.45\,GHz. (b) TE$_{10}$ attenuation versus frequency in copper waveguide.}
\label{fig:waveguide}
\end{figure}

  Fig.~\ref{fig:waveguide} presents the WR-340 waveguide analysis. Fig.~\ref{fig:waveguide}(a) shows the mode chart with cutoff frequencies for the five lowest-order modes. At 2.45\,GHz, only the TE$_{10}$ mode propagates, confirming single-mode operation. The single-mode bandwidth extends from 1.736\,GHz to 3.471\,GHz, providing substantial margin for magnetron frequency drift (typically $\pm$10\,MHz). Fig.~\ref{fig:waveguide}(b) shows the TE$_{10}$ attenuation in copper, which at 2.45\,GHz is only 0.009\,dB/m, resulting in negligible loss over the typical 1\,m waveguide run.

\subsection{Thermal management}

\begin{figure}[htb]
\centering
\includegraphics[width=0.95\textwidth]{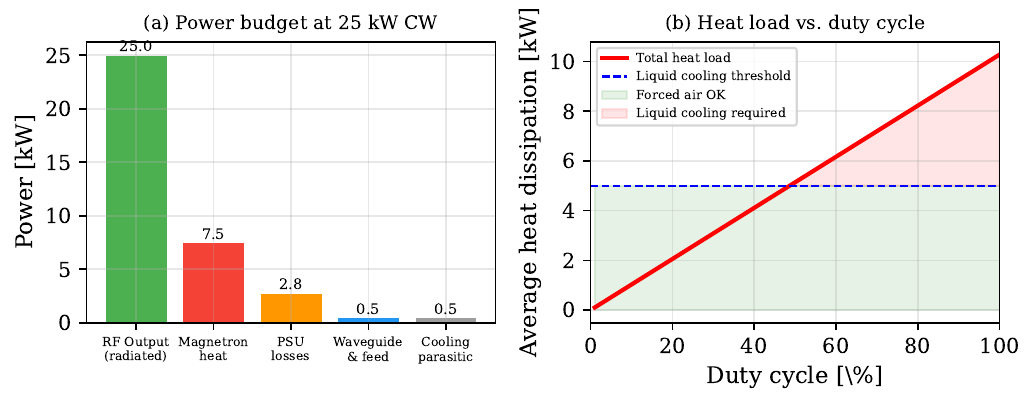}
\caption{Thermal analysis. (a) Power budget at 25\,kW CW, showing the RF output and heat dissipation in each subsystem. (b) Average heat load versus duty cycle; liquid cooling is required above the 5\,kW threshold.}
\label{fig:thermal}
\end{figure}

  Fig.~\ref{fig:thermal} presents the thermal analysis. At 25\,kW CW operation with 70\% magnetron efficiency, the system dissipates 7.5\,kW in the magnetron body and an additional 4.0\,kW in the PSU (90\% efficiency), for a total thermal load of approximately 12\,kW. Fig.~\ref{fig:thermal}(a) shows the power budget breakdown, and Fig.~\ref{fig:thermal}(b) illustrates the average heat load as a function of duty cycle. Liquid cooling is required above approximately 40\% duty cycle (5\,kW heat threshold), while pulsed operation at $\leq$5\% duty cycle permits forced-air cooling, significantly simplifying the system for mobile deployment.

\subsection{System efficiency chain}

\begin{figure}[htb]
\centering
\includegraphics[width=0.48\textwidth]{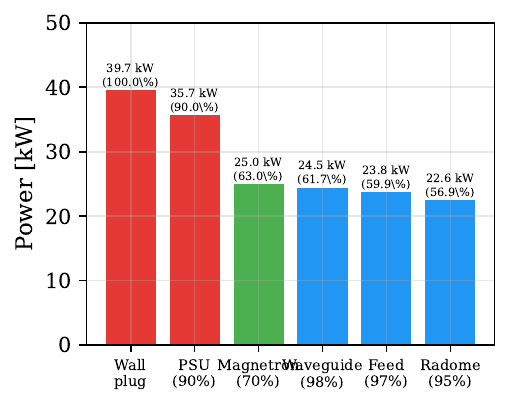}
\caption{Power flow through the system from wall-plug input to radiated RF, showing efficiency at each stage. Numbers indicate power level and cumulative efficiency.}
\label{fig:efficiency}
\end{figure}

  Fig.~\ref{fig:efficiency} shows the power flow from wall plug to radiated RF, illustrating the cumulative effect of losses at each stage. The overall wall-plug-to-radiated efficiency is approximately 58\%, consistent with the high DC-to-RF efficiency of cavity magnetrons~\cite{vyas2016efficiency,xi2025high,li2023particle,huang2025continuous}. The magnetron is the dominant loss mechanism (30\% of input power converted to heat), followed by the PSU (10\% losses).

\subsection{Human safety compliance}

\begin{figure}[htb]
\centering
\includegraphics[width=0.48\textwidth]{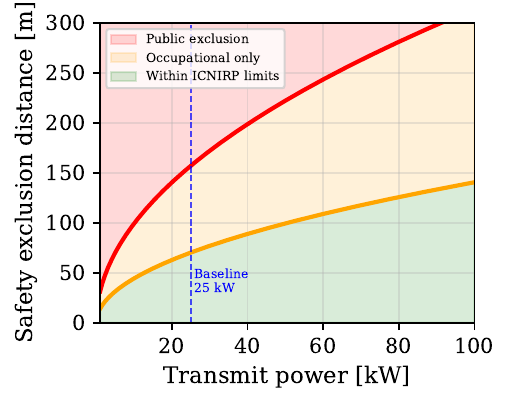}
\caption{ICNIRP safety exclusion distances versus transmit power for the 60\,cm dish. Green zone: within ICNIRP limits. Orange zone: occupational only. Red zone: general public exclusion.}
\label{fig:safety}
\end{figure}

  The ICNIRP 2020 guidelines~\cite{icnirp2020guidelines} specify reference levels for time-averaged exposure at 2.45\,GHz of 50\,W/m$^2$ (occupational) and 10\,W/m$^2$ (general public)~\cite{jeschke2022protection,heroux2023rf_limits}. Fig.~\ref{fig:safety} presents the calculated safety exclusion distances as a function of transmit power along the main beam axis. At the baseline 25\,kW CW, the occupational exclusion distance is 72\,m and the general public distance is 161\,m. These distances decrease with the inverse square root of power, and are proportionally smaller off-axis owing to the antenna pattern roll-off.

  In pulsed mode, the relevant exposure metric is the time-averaged power density, which for a 1\% duty cycle at 500\,kW peak power equals 5\,kW average---reducing the exclusion zones by a factor of $\sqrt{5}$ relative to 25\,kW CW. Implementation of ICNIRP-compliant exclusion zones requires integration with the safety interlock system, including personnel detection sensors and automatic beam inhibit.

\subsection{Dwell time analysis}

\begin{figure}[htb]
\centering
\includegraphics[width=0.95\textwidth]{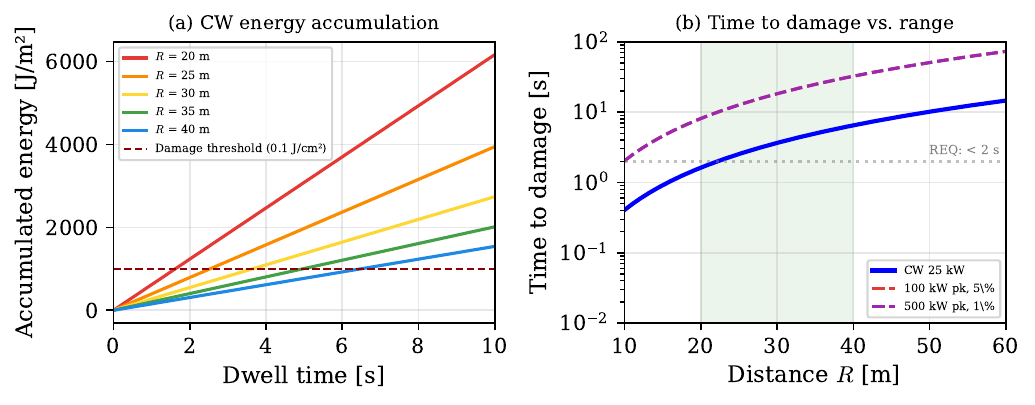}
\caption{Dwell time analysis. (a) CW energy accumulation versus time at five ranges. (b) Time to damage versus range for CW and pulsed modes.}
\label{fig:dwell}
\end{figure}

  Fig.~\ref{fig:dwell} presents the dwell time analysis, relating the required beam-on-target time to both range and operating mode. In CW mode, damage accumulates through thermal energy deposition; the time to reach the damage threshold (0.1\,J/cm$^2$) varies as $R^2$. At 20\,m, the dwell time is approximately 1.5\,s; at 40\,m, it increases to approximately 6\,s in CW mode. Pulsed operation at 500\,kW peak dramatically reduces the effective dwell time by inducing voltage-driven (non-thermal) breakdown on individual pulses, requiring only microseconds of cumulative exposure~\cite{huang2025failure,benford2024high}.

\section{Discussion}
\label{sec:discussion}

  The simulation framework presented in this paper provides several insights for HPM C-UAS system design that merit discussion.

  First, the Monte Carlo analysis demonstrates that deterministic link-budget calculations overestimate kill probability by approximately 60\% (Table~\ref{tab:mc_results}). This overestimation arises primarily from the polarisation mismatch and wire orientation randomness, which introduce a multiplicative loss factor that is unity only in the best case and averages to $\langle \cos^2\phi \rangle = 0.5$ over a uniformly distributed orientation. This finding underscores the importance of stochastic analysis for realistic C-UAS performance prediction and has implications for system specification: to guarantee 90\% kill probability at a given range, the system must be designed with 5--8\,dB margin above the deterministic threshold.

  Second, the pulsed operating mode provides the most effective path to extending engagement range within practical size, weight, and power (SWaP) constraints. At 1\% duty cycle, the 90\% kill range increases by approximately $4.7\times$ relative to CW operation at the same average power. This advantage is further amplified by the voltage-driven (rather than thermal) damage mechanism in pulsed mode, which requires only a few microsecond-duration pulses rather than seconds of continuous illumination~\cite{huang2023compact,he2025novel}.

  Third, the resonance-enhanced coupling model (Eq.~\ref{eq:resonance}) identifies drone wiring harnesses in the 5--8\,cm range as the critical vulnerability path at 2.45\,GHz. This finding is consistent with the recent experimental results of Zhang et al.~\cite{zhang2025investigation}, who reported that HPM irradiation caused GPS interference, datalink interruption, and ESC shutdown in consumer UAVs, with internal cable coupling identified as the primary energy entry mechanism.

  A comparison with fielded systems provides context for the simulation results. The Epirus Leonidas system uses solid-state GaN amplifier arrays rather than magnetrons, achieving software-defined beam control and demonstrated 100\% effectiveness against swarms of up to 49 drones. While the exact Leonidas operating parameters are not publicly disclosed, the system reportedly operates at power levels consistent with the pulsed configurations analysed in this study~\cite{nato2023cuas}. The Raytheon Phaser, which uses a traditional HPM source architecture more similar to the system modelled here, has also demonstrated operational effectiveness in field trials. The framework developed in this paper enables parametric comparison of such architectures within a unified analytical framework.

  Several limitations should be acknowledged. The damage probability model (Eq.~\ref{eq:sigmoid}) assumes far-field plane-wave illumination and does not capture near-field effects that become significant below approximately $2D^2/\lambda = 5.9$\,m for the 60\,cm dish. The sigmoid parameters in Table~\ref{tab:damage_thresholds} are derived from published literature spanning different experimental setups and device technologies; dedicated susceptibility testing of specific sUAS models would refine these values. The electromagnetic coupling model treats the wiring as an isolated short dipole and does not account for the complex electromagnetic environment inside a drone airframe, including mutual coupling between conductors and cavity resonances within the enclosure~\cite{gassab2023coupling}. Finally, atmospheric effects (humidity, rain) are neglected, which is reasonable at 2.45\,GHz where atmospheric attenuation is $<$0.01\,dB/km in clear air but may become significant in heavy precipitation.

\section{Conclusions}
\label{sec:conclusions}

  This paper has presented a multi-physics simulation framework for the design and performance evaluation of high-power microwave counter-UAS systems. The framework integrates RF propagation, antenna pattern analysis, electromagnetic coupling to unshielded drone electronics, and a probabilistic semiconductor damage model into a unified computational tool. The principal findings are as follows.

\begin{enumerate}[leftmargin=2em,itemsep=2pt]
  \item For a baseline 25\,kW CW system with a 60\,cm dish, the Monte Carlo simulation yields a kill probability of 51.4\% (95\% CI: [50.4, 52.3]\%) at 20\,m and 13.1\% (95\% CI: [12.4, 13.8]\%) at 40\,m, with the deterministic model overestimating performance by approximately $1.6\times$.

  \item Pulsed operation at 500\,kW peak power (1\% duty cycle) extends the 90\% kill range from approximately 18\,m to 88\,m while maintaining a manageable average power of 5\,kW, which permits forced-air rather than liquid cooling.

  \item The resonance-enhanced electromagnetic coupling model identifies drone wiring harnesses in the 5--8\,cm range (near $\lambda_0/2$ at 2.45\,GHz) as the critical vulnerability path, with induced voltages exceeding the 20--40\,V MOSFET gate oxide breakdown threshold at field strengths above 200\,V/m.

  \item The parametric design map reveals that the 90\% kill range scales as $\sqrt{P \cdot D^2}$, providing a practical tool for power--aperture trade studies.

  \item ICNIRP-compliant safety exclusion zones of 72\,m (occupational) and 161\,m (general public) are required at 25\,kW CW in the main beam, decreasing proportionally in pulsed mode.
\end{enumerate}

  Future work will extend the framework to include full-wave electromagnetic simulation of drone airframes using finite-difference time-domain (FDTD) methods, validation against controlled HPM exposure experiments on representative sUAS platforms, and integration of tracking system dynamics into the engagement probability model.

\section*{Acknowledgements}

The authors state that the style and English of this work has been polished using AI tools provided by \textit{QuillBot}. There is no funding associated with this work.

\bibliographystyle{unsrtnat}
\bibliography{ref}

\end{document}